\documentclass[useAMS,usenatbib,usegraphicx]{mn2e}

\def\ltsima{$\; \buildrel < \over \sim \;$}
\def\lsim{\lower.5ex\hbox{\ltsima}}
\def\gtsima{$\; \buildrel > \over \sim \;$}
\def\gsim{\lower.5ex\hbox{\gtsima}}

\newcommand{\be}{\begin{equation}}
\newcommand{\en}{\end{equation}}

\def\cmdue {\rm \ cm^{-2}}

\begin{document}

\title{The X--ray absorbing column densities of Swift Gamma--ray bursts} 
\author[S. Campana et al.]{S. Campana$^{1,}$\thanks{E-mail: sergio.campana@brera.inaf.it}, C. C. Th\"one$^{1}$,
A. de Ugarte Postigo$^1$, G. Tagliaferri $^1$, \newauthor  A. Moretti$^1$, S. Covino$^1$\\
$^1$ INAF-Osservatorio Astronomico di Brera, Via Bianchi 46, I--23807, Merate (Lc), Italy}

\maketitle

\begin{abstract}
Long gamma-ray bursts (GRBs) are associated with the explosion of
massive stars in star forming regions. A large fraction of GRBs show
intrinsic absorption as detected in optical spectra but absorption
signatures are also detectable in afterglow X--ray spectra. We present
here a comprehensive analysis the full sample of 93 GRBs with known
redshift promptly observed by Swift XRT up to June 2009. The
distribution of X--ray column densities clearly
shows that GRBs are heavily absorbed indicating that they indeed occur
in dense environments. Furthermore, there is a  lack of heavily
absorbed GRBs at low redshift ($z\lsim 1-2$) that might therefore be 
candidates for the missing `dark' GRB population. However, there is no
statistically significant correlation between the amount of X--ray
absorption and the `darkness' of a GRB.
Finally, we compare the hydrogen column densities derived in the
optical with those derived from X--ray absorption. The two
distributions are different, with the optical column densities being
lower than the X--ray ones, which is even more apparent when correcting for
metallicity effects. The most likely explanation is photoionization
of hydrogen in the circumburst material caused by the radiation field 
of the burst.
\end{abstract}


\section{Introduction}

(Long) gamma--Ray Bursts (GRBs) are recognized as being powerful cosmological tools, being detected 
from the local Universe to the highest redshifts reached today ($z\sim 8.2$)
(Salvaterra et al. 2009; Tanvir et al. 2009). GRBs are so bright that they can illuminate, for a short time, 
regions of distant galaxies otherwise inaccessible and probe the cosmic chemical evolution of the Universe 
(e.g. Fiore et al. 2005; Prochaska, Chen \& Bloom 2006; Fox et al. 2008). 
The GRB light shines through a medium that is very different (i.e. denser) from the typical
interstellar medium (ISM) observed with traditional tools in distant galaxies
High-resolution optical spectroscopic observations allow us to probe the relative abundances of different
elements highlighting the physical state of the gas, its chemical enrichment and dust content, and the
nucleosynthesis processes in the GRB star-forming environment (Savaglio 2006; Prochaska et al. 2007).
Despite the poor number statistics, a comparison between the metallicities obtained by
GRBs and those from studies of absorbing systems in the sightlines of quasar 
show that the former tend to be more metal-rich (on average by a
factor of 5 at high-$z$) and to have a flatter or no redshift evolution (see, e.g., Fynbo et al. 2008, 2009; 
Savaglio, Glazebrook \& LeBorgne 2009; Chen et al. 2009). 
Variations in the intensity of optical absorption lines were detected in a few cases of 
very optically bright GRBs. These
variations help to constrain the distance of the absorbing gas to the
GRB. This has been done in only two cases where distances larger than $\sim 2$ kpc have been derived. Matter
within this radius has been photoionized by the GRB radiation flux, which means that optical observations
can only probe the outermost part of the region surrounding the GRB (D'Elia et al. 2009; Vreeswijk et al. 2007).

Matter along the line of sight also attenuates X--ray photons. X--rays are photoelectrically absorbed by metals,
producing a well known and well identifiable pattern in the observed X--ray spectra (Morrison \& McCammon
1983). In contrast to absorption in the optical range, the absorption of X--rays is not sensitive to the state of 
the element (either gas or solid), therefore these
measurements provide an unbiased view of the total absorbing column density of material.
As demonstrated by several studies the X--ray absorbing column towards GRBs is high 
(Stratta et al. 2004; Campana et al. 2006; Watson et al. 2007), which implies in principle, i.e. for 
any known dust to gas ratio, a high dust extinction. 
In contrast, GRB optical afterglows, when present, often show a lack of reddening. Several suggestions have been advanced to 
explain this dichotomy but a clear understanding has not been reached yet (e.g. Galama \& Wijers 2001, see also Nardini 
et al. 2009). 

Stratta et al. (2004) presented a systematic analysis of a sample of 13 bright afterglows observed with the BeppoSAX
narrow field instruments. In only two cases they found a significant detection of additional intervening material in excess 
of the one in our Galaxy. However, due to the limited photon statistics, they could not exclude that intrinsic X--ray
absorption was also present in the other bursts. 
Campana et al. (2006) presented a systematic study of the intrinsic column density absorption in 6 GRBs with known 
redshift observed by Swift. Adding these GRBs  to others found in the literature with known redshift (mainly Chandra and 
XMM-Newton observations), they concluded that there is strong evidence that GRBs are intrinsically absorbed  
(Campana et al. 2006). The distribution of absorbing column densities is consistent with GRBs occurring within molecular clouds, 
which supports the link between (long) GRBs and star formation. Watson et al. (2007) found a large discrepancy between optical and 
X--ray derived GRB column densities and suggested photoionization of the gas cloud as 
the most likely cause.

Here we consider the full sample of 93 GRBs detected by Swift (Gehrels et al. 2004) XRT (Burrows et al. 2005) up to June 2009 
which were observed within 1,000 s (to ensure relatively high signal) from the burst onset and have a known redshift. In section 2 we 
briefly describe the data analysis and the way we evaluated the column density. In section 3 we discuss our results and in 
section 4 we draw our conclusions. 

\section{Data analysis}

We made extensive use of automated data products provided by the Swift/XRT GRB spectra repository (Evans et al. 2009).
For each burst all the data are processed in an homogeneous way and summed together. 
Most GRBs have one spectrum for each of the two instrument modes (Windowed Timing, WT, and 
Photon Counting, PC; Hill et al. 2004). For each time interval, a source and background spectrum, 
an exposure map and an ancillary response file (arf) are created and then combined together. 
To identify intervals affected by pile up we search for time intervals where the count rate
is above 0.6 counts s$^{-1}$ in PC mode or 150 counts s$^{-1}$ 
in WT mode. For those intervals  we then obtain the profile of the source and
compared it to the calibrated, non-piled-up, Point Spread Function (Moretti et al. 2005), 
for both PC and WT mode separately. Background spectra are created in an analogous way excluding 
contaminating sources in the background region. 
The individual arf files, which are generated in order to account for CCD defects, are combined using the {\tt addarf} tool,  
weighted by the counts in the individual source spectra. 

For each mode one source and one background spectrum are created. The source spectrum is binned using the {\tt grppha} 
tool such that there is at least one count in each spectral bin necessary for XSPEC to correctly calculate the Cash-statistic
(K. Arnaud, priv. communication) used to fit the spectra (Cash 1979). The spectra are modelled with an absorbed power 
law within XSPEC12.
The absorption model adopted is {\tt phabs}. The abundances are fixed at the solar value from Anders \&
Grevesse (1989). The photoelectric absorption consists of two components: one
from our Galaxy, kept constant to the value taken from Kalberla et al. (2005), and 
the other one free to vary, shifted in energy at the (fixed) redshift of the GRB.

\begin{figure}
\begin{center}
\includegraphics[width=6.0cm,angle=-90]{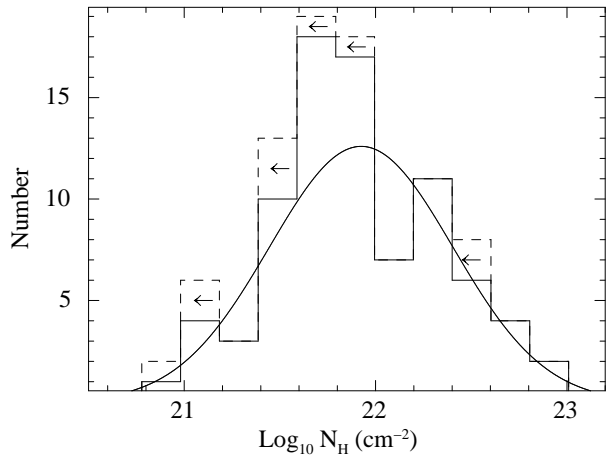}
\end{center}
\caption{Distribution of the X--ray column densities of Swift GRBs promptly observed by the XRT instrument. The 
continuous line histogram has been derived from positive detections, the dashed histogram also includes upper limits. The 
continuous line represents the fit with a lognormal function.}
\label{nh}
\end{figure}

Evans et al. (2009) provided already a collection of column densities of Swift GRBs. However,
either they did not fix the redshift in the fit of the X--ray column density and, when they fix it, they also included bursts without 
known redshift fixed at the mean value of the sample.
In addition, it has been shown that the absorbing column density can vary significantly when fitting 
an afterglow X--ray spectrum with a power law model when the overall spectrum is (slowly) changing (Butler \& Kocevski 2007; 
Falcone et al. 2006). In the case of a slowly varying spectrum a cut-off spectral model can be a solution to this problem.

For our sample, we selected GRBs with known spectroscopic redshift and with observations starting within 1,000 s from the burst onset.
The knowledge of the redshift is necessary in order to derive the intrinsic column density.
With these criteria our sample contains 93 GRBs up to May 2009. 
For these GRBs we derived the intrinsic absorbing column density selecting the time interval in the 
light curve when the 0.3-1.5 keV to 1.5--10 keV  hardness ratio is constant. During the first 
fast steep decay time interval the spectrum is strongly variable, as observed in several bursts. 
Therefore this interval is left out as well as any interval with (major) flaring 
activity. With these criteria we preferentially select observations in the $3,000-10^5$ s time interval and in PC mode,
after the early steep decay phase (Nousek et al. 2006).  
In a few cases when the late-time signal was not high enough to derive a constraint on the intrinsic column density,  
we selected short intervals in the early light curve during which the hardness ratio was constant (usually in WT mode). 
A comprehensive report of the selected time intervals and results is provided in Table 1.

\begin{table*}
\caption{Column densities for Swift GRBs with known redshift.}
\footnotesize{
\begin{tabular}{cccccccc}
GRB      & $z$   & $N_H(z)$             &$\Gamma$     & $N_H(\rm Gal)$  & Exp. time  & Time interv.       & Comments\\
         &       & ($10^{21}\cmdue$)    &             &($10^{20}\cmdue$)& (ks)       &     (s)            & \\
\hline
050126   &1.29	 &  $<1.2$              &$2.14\pm0.22$&    4.6	        &15.7        &  $100-10^5$        & PC - B11\\
050315   &1.949  &$9.4^{+1.1}_{-0.9}$   &$2.04\pm0.06$&    3.7          &108.0       &  $3000-10^6$       & PC \\
050319   &3.243  &$2.5^{+2.8}_{-2.3}$   &$1.96\pm0.08$&    1.3          &11.2        &  $5000-10^5$       & PC - B11\\
050401   &2.898  &$16.4^{+2.2}_{-2.2}$  &$1.85\pm0.05$&    4.4          &1.4         &  $200-2000$        & WT - D11\\
050416A  &0.654  &$6.1^{+0.5}_{-0.9}$   &$2.12\pm0.06$&    2.4          &92.1        &  $400-3\times 10^5$& PC - D3 - B11\\  
050525A  &0.606  &$1.5^{+0.9}_{-0.7}$   &$2.04\pm0.14$&    9.1          &5.8         &  $6000-3\times10^5$& PC - B11\\
050730   &3.969  &$12.0^{+7.7}_{-7.2}$  &$1.36\pm0.09$&    3.0          &0.1         &  $133-200$         & WT \\
050803   &0.422  &$2.0^{+0.2}_{-0.3}$   &$2.28\pm0.09$&    5.0          &187.0       &  $3000-10^6$       & PC - SM\\
050820A  &2.614  &$3.4^{+1.3}_{-1.3}$   &$1.98\pm0.04$&    4.4          &87.9        &  $3000-10^6$       & PC - D3 - B11\\
050826   &0.297  &$7.1^{+2.0}_{-2.0}$   &$2.59\pm0.19$&   19.0          &40.8        &  $3000-10^5$       & PC  - B11\\
050904   &6.29   &$63^{+34}_{-29}$      &$1.11\pm0.06$&    4.5          &0.1         &  $100-250$         & WT\\
050908   &3.347  &$17^{+11}_{-14}$      &$2.51\pm0.58$&    2.4          &49.9        &  $3000-3\times10^5$& PC - B11\\	
050922C  &2.200  &$5.2^{+1.7}_{-1.9}$   &$2.32\pm0.10$&    5.4          &81.2        &  $300-3\times 10^5$& PC - B11\\
051016B  &0.936  &$7.7^{+1.1}_{-1.0}$   &$2.04\pm0.12$&    3.2          &130.2       &  $3000-10^6$       & PC - B11\\
051109A  &2.346  &$5.0^{+3.0}_{-2.7}$   &$2.03\pm0.06$&   16.0          &10.0        &  $3000-10^6$       & PC - B11\\
060115   &3.533  &$22.4^{+9.3}_{-8.5}$  &$1.84\pm0.12$&    9.5          &0.1         &  $100-160$         & WT - B11\\
060124   &2.300  &$7.3^{+0.9}_{-1.1}$   &$2.06\pm0.04$&    9.0          &162.9       &  $110-10^6$        & PC - B11\\
060206   &4.056  &$12.7^{+8.0}_{-11.8}$ &$2.20\pm0.21$&    0.9          &106.9       &  $700-10^5$        & PC - B11\\
060210   &3.912  &$18.9^{+3.4}_{-3.3}$  &$2.09\pm0.04$&    6.1          &76.9        &  $3000-10^6$       & PC - D3 - D11\\
060218   &0.033  &$5.2^{+0.5}_{-0.5}$   &$1.45\pm0.11$&    9.4              &9.4      &  $100-200$ &       WT - B11\\
060223A  &4.41   & $<37.6$              &$2.12\pm0.26$&    6.9          &18.4        &$10^2-10^3$ \& $4000-10^5$& PC - B11\\
060418   &1.489  &$3.4^{+1.1}_{-1.0}$   &$1.99\pm0.08$&    8.8          &0.4         &  $250-1000$        & WT - B11\\
060502A  &1.510  &$4.7^{+0.9}_{-1.0}$   &$2.03\pm0.11$&    3.5          &89.1        &  $3000-10^6$       & PC - B3 - B11\\
060510B  &4.9    &$44.7^{+9.7}_{-9.1}$  &$1.06\pm0.04$&    4.1          &0.1         &  $100-200$         & WT - SM - D3 - B11\\
060512   &2.1    & $<1.4 $              &$2.09\pm0.20$&    1.5          &83.2        &  $3000-3\times10^5$& PC - B11\\
060522   &5.11   &$34^{+25}_{-13}$      &$2.29\pm0.18$&    4.1          &17.6        &  $300-10^5$        & PC - B11\\
060526   &3.221   & $<9.8$               &$2.03\pm0.32$&    5.0          &39.2        &  $80-120$\& $450-10^3$& WT - B11\\  
060605   &3.78   &$7.8^{+3.7}_{-2.6}$   &$2.18\pm0.08$&    3.9          &37.2        &  $3000-3\times10^5$& PC - B11\\
060607A  &3.075  &$4.8^{+2.5}_{-2.4}$   &$1.66\pm0.06$&    2.4          &27.3        &  $600-10^5$        & PC - B11\\
060707   &3.424  &$<5.4$                &$2.07\pm0.13$&    1.4          &3.1         &  $180-2\times 10^6$& PC - B11\\
060714   &2.711  &$31.1^{+7.1}_{-8.4}$  &$2.22\pm0.10$&    6.1          &182.3       &  $300-3\times 10^6$& PC - B11\\
060729   &0.543  &$1.4^{+0.2}_{-0.2}$   &$2.08\pm0.04$&    4.5          &69.5        &  $3000-3\times10^5$& PC - B11\\
060904B  &0.703  &$4.4^{+0.7}_{-1.2}$   &$2.21\pm0.14$&   11.8          &69.9        &  $3000-3\times10^5$&PC - B11\\
060906   &3.686  &$22.5^{+9.9}_{-12.8}$ &$2.11\pm0.16$&    9.8          &21.3        &  $1000-10^5$       & PC - VAR - B3 - B11\\
060908   &1.884  &$5.9^{+2.6}_{-1.5}$   &$2.11\pm0.17$&    2.3          &11.6        &  $200-10^5$        & PC - B3 - B11\\
060926   &3.209  &$34^{+23}_{-20}$      &$2.04\pm0.28$&    7.6          &10.1        &  $3000-10^5$       & PC - B11\\
060927   &5.464  & $<39$                &$1.96\pm0.18$&    4.6          &8.6          &  $100-10^5$       & PC \\
061007   &1.262  &$4.5^{+0.3}_{-0.3}$   &$1.88\pm0.02$&    1.8          &1.9          &  $90-2000$        &  WT - B11\\
061110A  &0.758  &$1.1^{+1.2}_{-0.6}$   &$1.90\pm0.27$&    4.3          &238.0        &  $700-10^6$       & PC - B11\\
061121   &1.315  &$5.4^{+0.5}_{-0.6}$   &$1.90\pm0.05$&    4.0          &120.5        &  $600-10^6$       & PC \\
061222B  &3.355  &$43^{+42}_{-30}$      &$2.96\pm0.51$&   27.7          &0.1          &  $193-212$        &  WT - B11\\
070110   &2.352  & $<3.8$               &$2.21\pm0.13$&    1.6          &2.4          &  $3000-8000$      & PC - B11\\
070208   &1.165  &$8.6^{+3.0}_{-2.6}$   &$2.17\pm0.30$&    1.8          &53.4         &  $3000-10^6$      & PC - B3 - B11\\
070306   &1.497  &$21.9^{+3.6}_{-3.3}$  &$1.92\pm0.11$&    2.9          &6.0          &  $10^4-4\times10^4$& PC - D11\\
070318   &0.840  &$7.1^{+0.7}_{-1.3}$   &$2.19\pm0.12$&    1.4          &129.4        &  $6000-10^6$      & PC - B11\\
070411   &2.954  &$16^{+16}_{-13}$      &$2.29\pm0.21$&   28.7          &4.3          &  $450-20000$      & PC - B11\\
070419A  &0.971  &$3.5^{+1.1}_{-1.0}$   &$2.32\pm0.16$&    2.4          &0.1          &  $200-300$        & WT - B3 - B11\\
070506   &2.309  &$3.0^{+5.0}_{-2.7}$   &$2.04\pm0.24$&    3.8          &2.7          &  $400-8000$       & PC - VAR - B11\\
070529   &2.50   &$62^{+22}_{-12}$      &$2.33\pm0.19$&   19.3          &120.4        &  $140-10^6$       & PC - B11\\
070721B  &3.630  &$12.9^{+7.4}_{-4.3}$  &$1.67\pm0.06$&    2.3          &25.6         &  $3000-10^5$      & PC\\
070802   &2.454  &$12.1^{+8.1}_{-9.9}$  &$2.08\pm0.17$&    2.9          &12.1         &  $140-58000$      & PC - D11\\
070810A  &2.17   &$6.8^{+2.8}_{-1.5}$   &$2.14\pm0.13$&    1.8          &15.9         &  $100-40000$      & PC - B11\\
071020   &2.146  &$5.2^{+1.8}_{-1.7}$   &$1.84\pm$0.07&    5.1          &0.2          &  $70-300$         & WT - B3 - B11\\
071031   &2.692  &$9.6^{+2.3}_{-2.2}$   &$2.39\pm0.09$&    1.2          &0.1          &  $250-350$        & WT - B11\\
071112C  &0.823  &$0.6^{+0.5}_{-0.5}$   &$1.73\pm0.07$&    7.4          &0.3          &  $90-400$         & WT - B11\\
071122   &1.14   &$2.2^{+1.2}_{-1.2}$   &$2.10\pm0.16$&    4.8          & 0.1         &  $150-300$        & WT - B3 - B11\\
\hline
\end{tabular}
}
\end{table*}

\setcounter{table}{0}
\begin{table*}
\caption{continued.}
\footnotesize{
\begin{tabular}{cccccccc}
GRB      & $z$   & $N_H(z)$             &$\Gamma$     & $N_H(\rm Gal)$   & Exp. time    & Time interv.      & Comments\\
         &       & ($10^{21}\cmdue$)    &             &($10^{20}\cmdue$) & (ks)         &     (s)           & \\
\hline
080210   &2.642  &$15.9^{+6.1}_{-7.8}$  &$2.24\pm0.11$&    5.5           &30.0          &  $250-10^5$       & PC\\
080310   &2.427  &$3.8^{+1.8}_{-2.2}$   &$1.98\pm0.10$&    3.3           &103.9         &  $3000-10^6$      & PC - B3\\
080319B  &0.938  &$1.3^{+0.1}_{-0.1}$   &$1.78\pm0.02$&    1.1           &0.92          &  $800-2000$       &  WT - B3\\
080319C  &1.949  &$7.5^{+2.0}_{-1.8}$   &$1.70\pm0.08$&    2.2           &23.8          &  $200-3\times10^5$& PC - D3\\
080330   &1.512  & $<2.6$               &$1.89\pm0.13$&    1.2           &1.4           &  $100-2000$       & PC\\
080411   &1.030  &$4.6^{+0.4}_{-0.4}$   &$2.03\pm0.04$&    5.8           &69.3          &  $200-10^6$       & PC\\
080413A  &2.433  &$16.8^{+5.1}_{-4.6}$  &$3.39\pm0.26$&    8.7           &0.1           &  $80-150$         & WT\\
080413B  &1.101  &$2.7^{+0.5}_{-0.5}$   &$1.99\pm0.06$&    3.1           &50.9          &  $500-3\times10^5$&PC\\
080430   &0.767  &$3.9^{+0.3}_{-0.3}$   &$2.12\pm0.07$&    1.0           &117.9         &  $3000-10^6$      & PC\\
080520   &1.546  &$12.9^{+6.4}_{-5.6}$  &$2.32\pm0.23$&    6.8           &4.6           &  $100-2\times10^4$& PC\\
080603B  &2.689  &$7.6^{+3.1}_{-2.9}$   &$1.84\pm0.10$&    1.2           &0.17          &  $100-300$        & WT\\
080604   &1.417  & $<0.71$              &$2.19\pm0.08$&    3.8           &0.1           &  $100-220$        & WT\\
080605   &1.640  &$7.2^{+0.8}_{-0.8}$   &$1.76\pm0.04$&    6.7           &0.6           &  $100-800$        & WT\\
080607   &3.037  &$36.0^{+3.3}_{-3.2}$  &$2.20\pm0.06$&    1.7           &0.36          &  $200-600$        & WT\\
080707   &1.232  &$3.3^{+1.9}_{-1.9}$   &$2.08\pm0.17$&    7.0           &20.9          &  $100-10^5$       & PC\\
080721   &2.591  &$7.1^{+0.6}_{-0.6}$   &$1.83\pm0.02$&    6.9           &1.3           &  $100-2000$       & WT\\
080804   &2.205  &$2.4^{+1.7}_{-1.5}$   &$1.86\pm0.08$&    1.6           &40.4          &  $200-3\times10^5$& PC\\
080805   &1.504  &$19.5^{+7.1}_{-7.1}$  &$2.47\pm0.15$&    3.5           &89.9          &  $300-10^6$       & PC\\
080810   &3.360  &$4.3^{+2.9}_{-2.3}$   &$2.16\pm0.08$&    3.3           &26.9          &  $3000-2\times10^5$& PC\\
080905B  &2.374  &$31.1^{+3.4}_{-5.9}$  &$2.13\pm0.12$&    3.5           &99.6          &  $500-10^6$       & PC\\
080913   &6.7    &$95^{+89}_{-77}$      &$2.00\pm0.24$&    3.2           &166.9         &  $100-10^6$       & PC - VAR\\
080916   &0.689  &$9.0^{+2.0}_{-3.5}$   &$2.43\pm0.26$&    1.8           &109.7         &  $2\times10^4-10^6$& PC\\
080928   &1.692  &$3.1^{+1.1}_{-1.0}$   &$2.13\pm0.07$&    5.6           &27.6          &  $3000-10^5$      & PC\\
081007   &0.530  &$6.2^{+0.5}_{-0.5}$   &$2.20\pm0.09$&    1.4           &102.5         &  $200-10^6$       & PC\\
081008   &1.969  &$4.6^{+1.7}_{-2.1}$   &$2.13\pm0.11$&    7.1           &30.9          &  $500-3\times10^5$& PC\\
081028   &3.038  &$6.4^{+1.6}_{-2.0}$   &$2.11\pm0.05$&    4.0           &176.2         &  $10^4-10^6$      & PC\\
081118   &2.58   & $<3.8$               &$2.22\pm0.24$&    3.7           &99.4          &  $200-10^6$       & PC\\
081203A  &2.05   &$3.1^{+1.7}_{-1.7}$  &$1.73\pm0.09$&    1.7           &0.1           &  $200-300$        &  WT\\
081222   &2.77   &$4.6^{+1.2}_{-1.1}$   &$2.01\pm0.04$&    2.2           &0.8           &  $60-1000$        &  WT\\
090102   &1.547  &$6.8^{+0.7}_{-1.1}$   &$1.88\pm0.07$&    4.1           &55.4          &  $700-3\times10^5$&PC\\
090205   &4.650  &$10.4^{+9.8}_{-8.5}$  &$2.11\pm0.13$&    7.7           &48.           &  $100-3\times10^5$&PC - VAR\\
090418A  &1.608  &$11.7^{+2.0}_{-1.8}$  &$2.09\pm0.09$&    3.6           &8.4           &  $400-2\times10^4$& PC\\
090423   &8.1    &$102^{+49}_{-54}$     &$1.98\pm0.15$&    2.9           &36.7          &  $3000-3\times10^5$& PC - VAR\\
090424   &0.544  &$5.1^{+0.4}_{-0.3}$   &$2.09\pm0.07$&    1.9           &213.1         &  $2000-3\times10^6$& PC\\
090516   &4.109  &$22.9^{+4.0}_{-3.9}$  &$2.17\pm0.06$&    4.5           &73.6          &  $3000-3\times10^5$& PC\\
090519   &3.85   &$26^{+19}_{-19}$      &$1.57\pm0.20$&    3.0           &41.9          &  $200-3\times10^4$& PC - VAR\\ 
090529   &2.625  &$4.3^{+2.2}_{-2.1}$   &$2.71\pm0.13$&    1.6           &0.2           & $200-400$         & WT - VAR\\ 
\hline
050505   &4.27   &$18.7^{+2.5}_{-2.9}$  &$2.04\pm0.05$&    1.7           &104.8         & $3000-10^6$       & PC \\
061110B  &3.434  &$31^{+28}_{-22}$      &$2.21\pm0.32$&    3.4           &29.8          & $3000-10^5$       & PC - VAR\\ 
070611   &2.039  & $<5.6$                &$1.92\pm0.24$&    1.4           &17.1          & $3000-5\times10^4$& PC \\ 
\hline
\end{tabular}
}

\noindent
Errors at $90\%$ confidence level. Upper limits are at $3\,\sigma$ level. For the error on the power law slope we just quote 
the maximum between lower and upper values.

\noindent 
PC and WT refer to the Swift XRT observing mode. SM or VAR is a small or variable hardness ratio. D3 and B3 refer
to burst which are optically dark or
bright at 1,000 s. Data are taken from Cenko et al. (2009). D3 are bursts with the optical ($R$) to X--ray spectral index $\beta_{OX}$ evaluated at 1,000 s lower than 0.5 (Jakobsson et al. 2004). B3 bursts have $\beta_{OX}>0.5$.
D11 and B11 are dark and bright bursts from the sample of Zheng et al. (2009) where $\beta_{OX}$ has been evaluated at 11 hr.
B11 are bursts with we took $\beta_{OX}>0.6$.  Note that some GRBs that are `dark' at 1,000 s are considered `bright' at 11 hr.

\noindent 
The last three GRBs in the table have XRT observations starting later than 1,000 s. We list them here 
because they have an optically determined hydrogen column density.

\end{table*}

\begin{figure}
\begin{center}
\includegraphics[width=6.0cm,angle=-90]{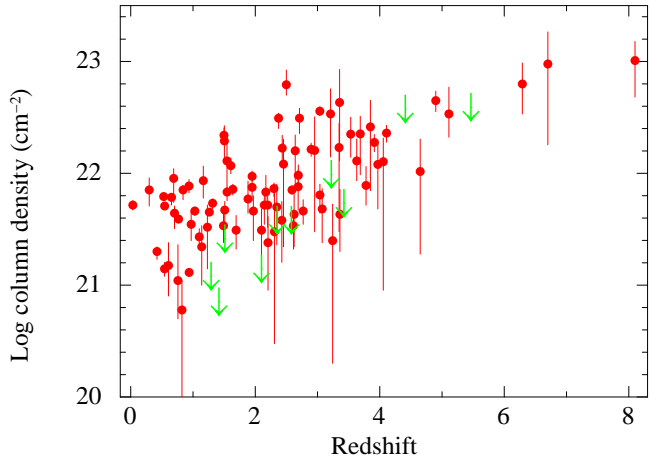}
\end{center}
\caption{X--ray column densities versus redshift. Upper limits also plotted as arrows.}
\label{nh2}
\end{figure}

\section{Results and Discussion}

We analyzed the XRT spectra of 93 GRBs promptly observed by Swift with known redshift (up to May 2009). 
For only 8 of them we do not find evidence of intrinsic X--ray absorption at the host galaxy site.
This clearly indicates that the medium surrounding GRBs is not tenuous.  In contrast to optical studies which are subject 
to uncertainties due to ionizations level and elements depleted to dust, X--ray absorption studies are less biased since the absorption is 
produced by the innermost electrons and therefore does not depend on the state of the absorbing medium.  
The distribution of intrinsic absorbing column densities is shown in Fig. 1. This distribution can be well 
described by a lognormal distribution with mean $\log N_H(z)=21.9\pm0.1$ ($90\%$ confidence level) and standard deviation
$\sigma_{\log N_H(z)}=0.5\pm0.1$ (the fit is good with a reduced $\chi^2$ of 1.2). Given the size of our sample this
clearly indicates that X--ray afterglow spectra are intrinsically absorbed (see also Nardini et al. 2009). 
The amount of this absorption is typical of the plane of our Galaxy considering solar metallicity. 

In order to determine the origin of the absorption excess, we compare the distribution of measured 
intrinsic column densities with the distribution expected for bursts occurring in different ambient media. We already demonstrated
(Campana et al. 2006) that the column density distribution is consistent with Galactic-like molecular clouds 
(Reichart \& Price 2002) and not with a random GRB occurrence within a spiral galaxy like our own (Vergani et al. 2004).
Our extended sample strengthens this conclusion since a random occurrence in our Galaxy would predict a 
sizable ($\sim 30\%$) GRB population with no intrinsic absorption which is not present in our sample.
Following Reichart \& Price (2002) we parametrize the column density distribution with a broken power-law
\begin{equation}
n(\log{N_H}) = \cases{ 
\frac{1}{\ln{10}}\frac{bc}{c-b}\left(\frac{N_H}{a}\right)^b & ($0 < N_H < a$) \cr 
\frac{1}{\ln{10}}\frac{bc}{c-b}\left(\frac{N_H}{a}\right)^c & ($N_H \ge a$)},
\label{model}
\end{equation}
where $a$ is the break column density, $b$ is the low column density power-law index, and 
$c$ is the high column density power-law index. 
We find that $\log{(a/{\rm cm}^{-2})} = 21.71^{+0.14}_{-0.15}$, $b=1.59^{+1.81}_{-0.57}$, and 
$c=-0.78^{+0.42}_{-0.26}$ ($90\%$ confidence level).
We stress that our sample also contains upper limits and that
we are sensitive to low values of column density, which however
are found only in a small fraction of the total sample.
If we include 6 additional tight upper limits (limits with $N_H<4\times10^{21}\cmdue$) we obtain very similar values 
as expected because these additional GRBs comprise just $7\%$ of the sample.
These results strongly support an origin of long GRBs within high density regions of host galaxies.

\begin{figure}
\begin{center}
\includegraphics[width=6.0cm,angle=-90]{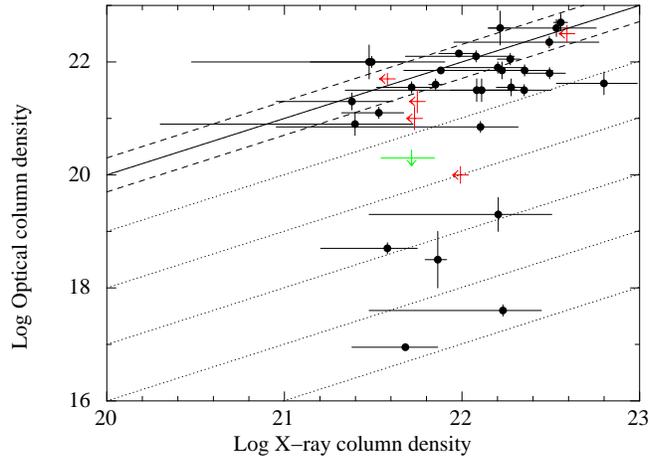}
\end{center}
\caption{X--ray column densities versus optically derived column densities along with their errors. Dashed lines 
indicate values within a factor of two from the line of equal X--ray to column density (upper continuous line). Dotted lines 
mark optical column densities $n$th orders of magnitude less than X--ray ones.}
\label{nh3}
\end{figure}

With our large sample we can also check if there is any dependence on redshift. To detect an intrinsic column density, 
one preferentially selects higher column density at higher redshifts since the column density contribution is shifted 
towards lower energy bands and therefore more difficult to identify. 
To get the same absorption contribution in an X--ray spectrum at different redshift, the intrinsic column density 
should increase as $\sim (1+z)^{2.6-2.7}$ (e.g. Galama \& Wijers 2001). 
The interesting point is that of the ten GRBs at redshift $z\gsim4$ only two have upper limits (see Fig. 2).
The mean value of this absorption (excluding upper limits) is $\sim 5\times 10^{22}\cmdue$ with a large dispersion.
This indicates that the available space for a weakly absorbed GRBs is at a level of only $\sim 20\%$. To quantify this
statement we compare the column density distribution of bursts at $z>4$ with the ones at $z<1$. For the two upper limits in 
the $z>4$ sample we take a worst case approach an fix them to the mean column density of the $z<1$ sample 
(i.e. $5\times10^{21}\cmdue$, see also below).
A Kolmogorov-Smirnov (KS) test of the two distributions shows that they are not drawn from the same parent distribution with a
probability of $0.08\%$ (equivalent to Gaussian $3.3\,\sigma$). This value does not change for lower values of the high-$z$ 
upper limits.
This implies either the lack of low absorption burst at high redshift or the lack of heavily absorbed bursts at low redshift. 
The first instance might be explained by a higher mean and/or compactness of star forming regions at high redshifts.
Alternatively, in a few cases, absorption caused by intervening systems can also play a role (Campana et al. 2006). 
The second instance might find an explanation in the change of optical extinction curve (e.g. due to a change in dust composition, 
dust to gas ratio, grain size): at high redshifts the extinction curve should be much flatter and similar to the one determined 
for quasars (Maiolino et al. 2004), at lower redshifts the curve should resamble more closely the one of our Galaxy, with a higher
optical absorption for the same X--ray absorbing column density. If this is true heavily absorbed bursts at low redshift are missing 
from our sample because for these bursts we cannot get an optical counterpart and/or a redshift, i.e. they likely populate 
the `dark' burst class. 

We therefore search for a possible correlation between the presence of high X--ray 
absorbing column density and the darkness of a GRB. Following Jakobsson et al. (2004) a GRB is classified as dark if the 
optical ($R$ band) to X--ray spectral index $\beta_{OX}$ is smaller than 0.5. 
We first consider the sample of Cenko et al. (2009) where $\beta_{OX}$ has been evaluated at 1,000 s after the burst 
detection (see also Perley et al. 2009).
In this sample there are 5 dark bursts and 9 non-dark bursts common to our sample. 
A KS test on the distribution of column densities of the two sample provides a probability of $15\%$ of being drawn 
from the same population. We also consider the work by Zheng et al. (2009) in which $\beta_{OX}$ has been evaluated at 
11 hr (as the original definition by Jakobsson et al. 2004). The two optical samples are not overlapping and some 
burst classified as `dark' in one sample is not dark in the other and viceversa.
In this case we have 4 dark GRBs and 39 firm non-dark GRBs 
(with $\beta_{OX}>0.6$) in common with our sample. We included in the non-dark burst sample three upper limits (with $N_H<
4\times 10^{21}\cmdue$) taking their upper limit as the true value. In this case a KS test provides a probability of $2\%$ 
(corresponding to a $2.3\,\sigma$ Gaussian probability). 
This might suggest a different population but  the dark population lack a sufficiently high number of GRBs 
to perform a reasonable statistical analysis.

It has been suggested that not only intrinsic absorption might be a cause of darkness of a GRB but also the 
redshift.  We therefore tried a two parameters (or bi-dimensional) KS test (Press, Flannery \& Teukolsky 1986), 
testing together the X--ray column density and the redshift.
In the case of $\beta_{OX}$ evaluated at 1,000 s, the 2D-KS probability is $16\%$,  in the case of $\beta_{OX}$ 
evaluated at 11 hr the 2D-KS probability is $16\%$.
This indicates that X--ray absorption alone or together with redshift cannot provide a statistically sound explanation for the darkness 
of GRBs. Cenko et al. (2009) suggested that optical absorption is the likely cause, but based on their sample a KS test 
of the host absorption of dark and non-dark GRBs cannot statistically support this suggestion. 

Last, we compare X--ray column densities to the hydrogen column densities derived from optical data (available for GRBs 
at $z\gsim 2$). This is now possible thanks 
to our large X--ray sample and the optical GRB sample recently presented by Fynbo et al. (2009). 
These two samples comprise 36 common GRBs (including three more from the literature, GRB050505, GRB050904, 
and GRB081203). 
Out of these 36 GRBs, 5 GRBs have only an upper limit in X--rays and one has an upper limit in the optical (see Fig. 3).
A KS test of the 30 GRB X--ray and optical column densities provide a probability of $1.1\%$ (Gaussian $2.5\,\sigma$)
that the two populations come from the same distribution. As we see in Fig. 3, the bulk of the GRBs have column densities
in the X--rays which are within a factor of $\sim 10$ larger than the ones in the optical. In addition, none of the bursts has an optical 
column density larger than a factor of $\sim 2$ compared to the X--ray one, including errors, which indicates that X--ray 
column densities are, as expected, good tracers of the ongoing absorption. This is not the case for optical column
densities since a large number of them lies well below the X--ray derived value. 
The most extreme GRBs have ratio of X--ray to optical column density in the 
$10^{-5}-10^{-4}$ range. Several possibilities have been discussed in the literature. For differences of an order of magnitude, higher
metallicities might be invoked, providing more metals (X--ray absorption) at the same value of hydrogen (Campana et al. 2007, 2008; 
Watson et al. 2007).  For larger differences the likely explanation relies on ionization of hydrogen by the GRB high energy flux 
(Lazzati, Perna \& Ghisellini 2001; Watson et al. 2007) in the close GRB environment. 

 From spectroscopic studies of quasar Damped Lyman Alpha (DLA) we know that the Si\,II $\lambda$1526 line intensity 
can be used as a proxy for metallicity $M$ through the relation $[M/H] = -0.92 + 1.41\, \log (EW/{\rm\AA})$ 
(with $EW$ being the equivalent width of the SiII line; Prochaska et al. 2008). 
GRBs seem to follow the same correlation, although possibly with a somewhat steeper slope, but the sample of bursts with 
optically derived metallicities and the detection of the Si II $\lambda$1526 line is still very limited. 
Following this suggestion we derived a rough metallicity value for 29 GRBs based on the Si\,II line strength. 
For 24 of them (including 3 upper limits) we have also a determination of the optical column density.
With this sample we first investigate if there is a correlation between the derived metallicity and the X--ray column densities 
and find that there is none.
As a further step we assume that the metallicity derived from the SiII lines applies also to the medium absorbing the X--ray flux.
If this is the case the true column density absorbing the X--ray flux is just the derived value divided by the metallicity value,
since the X--ray fitting only allows to determine the integrated contribution of $N_H/M$. 
We can compare in Fig. 4 the `corrected' X--ray column densities with the optically derived ones for 24 GRBs. 
In this case no metallicity effect should be present. The disagreement between the two is even 
more severe. The probability that the distribution of the optical and X--ray column densities were drawn from the same parent 
population is $8\times 10^{-7}$, equivalent to Gaussian $4.9\,\sigma$, which calls for ionization of hydrogen by the GRB 
high energy flux.

\begin{figure}
\begin{center}
\includegraphics[width=6.0cm,angle=-90]{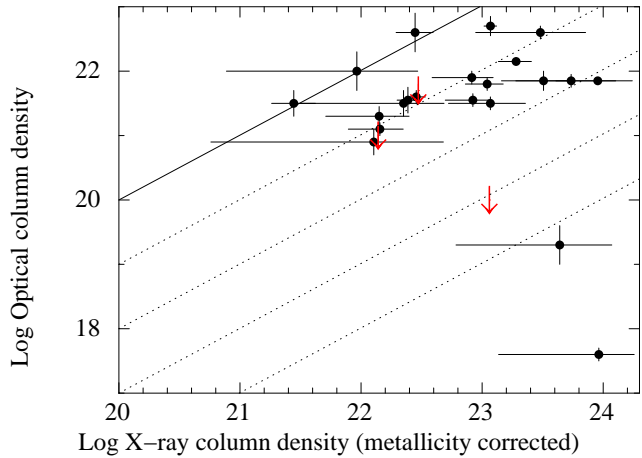}
\end{center}
\caption{X--ray versus optical column densities. X--ray column densities were corrected for metallicities different 
from the solar ones adopting the ones measured in the optical. The continuous line refers to equal optical to X--ray values,
the dotted lines to $n$th orders of magnitude difference.}
\label{nh4}
\end{figure}

\section{Conclusions}

We considered a complete sample of 93 Swift long GRBs with prompt XRT pointing ($\lsim 1,000$ s) 
and with a spectroscopic redshift. The distribution of the intrinsic X--ray absorption column density is consistent with a 
lognormal distribution with mean $\log N_H(z)=21.9\pm0.1$ ($90\%$ confidence level) and standard deviation
$\sigma_{\log N_H(z)}=0.5\pm0.1$. This confirms that long GRBs occur in a dense environment, which is consistent with 
the expected distribution of absorptions for GRBs occurring randomly in molecular clouds (and not with long GRBs 
occurring randomly in any place of a galaxy similar to own Galaxy).

Looking at the distribution of X--ray column densities versus redshift we find a lack of non-absorbed GRBs at high redshift and 
a lack of heavily absorbed GRBs at low redshift. The former might be explained in terms of more compact and dense star formation
regions in the young Universe (or to a sizable contribution from intervening systems). The latter might be interpreted as due to 
a change in the dust properties with redshift, with GRBs at redshift $z\lsim 2-3$ having a higher dust to gas ratio for 
the same X--ray column density. This will naturally provide a lack of heavily (X--ray) absorbed GRBs at small redshifts, 
i.e. dark GRBs.
We also investigated in more detail the relation between X--ray absorption and darkness of a GRB. We considered two different 
samples of long GRBs selected at different times (1,000 s and 11 hr). For none of the two we found a statistically significant relation 
between X--ray absorption and darkness, nor in a two-dimensional test including X--ray absorption and redshift. We note however 
that the sample of dark GRBs with known redshift is small and the weakness of our correlation might also depend on this.

Finally, we compare the hydrogen column densities derived in the optical and those from X--ray absorption. The two distributions 
are markedly different, with the optical column densities being lower than X--ray ones (and even more if correcting for metallicity effects). 
The most likely explanation is in terms of photoionization (depletion) of hydrogen in the circumburst material caused by the burst 
radiation field (Lazzati et al. 2001;  Watson et al. 2007).

\section*{Acknowledgments}
This work is supported at OAB-INAF by the ASI grant I/011/07/0. This work made use of data supplied by the UK Swift Science 
Data Centre at the University of Leicester. We thank Phil Evans for useful suggestions and for keeping updated the Swift site.

\end{document}